\documentclass[preprint,12pt]{article}

\usepackage{amssymb}
\usepackage{amsthm}

\usepackage{latexsym}
\usepackage{enumerate}
\usepackage{amsmath, amsfonts}
\usepackage{graphicx}

\renewcommand{\Vec}[1]{\mbox{\boldmath $#1$}}
\newcommand{\Real}{\mathbb{R}}
\newcommand{\Int}{\mathbb{Z}}
\newcommand{\hypn}{\mathchar`-}
\newcommand{\sumW}{W_+}
\newcommand{\sumWk}[1]{W_{#1+}}
\newcommand{\tT}{t^{\bullet}}
\newcommand{\tArc}{A^{\bullet}}
\newcommand{\tGamma}{\Gamma^{\bullet}}

\begin{document}

\title{Pseudo Polynomial Size LP Formulation  \\
for Calculating the Least Core Value \\
 of Weighted Voting Games
\thanks{A preliminary version of this paper was presented at 
the 12th Annual Meeting of the Asian Association
 for Algorithms and Computation,
 April 19 - 21, 2019, Seoul, South Korea. 
This work was supported by JSPS KAKENHI Grant Numbers 
JP26285045, JP26242027, JP20K04973. 
We thank Kento Tanaka for extensive discussions.}
}

\author{
\begin{tabular}{lll@{}ll}
&Tokyo Institute of Technology&& Masato Tanaka\\
&Tokyo Institute of Technology&& Tomomi Matsui
\end{tabular}
}

\date{\today}

\maketitle

\begin{abstract}
In this paper, we propose 
	a pseudo polynomial size LP formulation
	for finding a payoff vector in the least core 
	of a weighted voting game.
The numbers of variables and constraints
	in our formulation are both bounded by $\mbox{O}(n \sumW)$, 
	where $n$ is the number of players and 
	 $\sumW$ is the total sum of (integer) voting weights.  
When we employ our formulation, 
	a commercial LP solver calculates a payoff vector
	in the least core of practical weighted voting games in a few seconds.
We also extend our approach to vector weighted voting games.
\end{abstract}


\section{Introduction}
\label{section:introduction}

This paper deals with a special class of simple games, 
	called {\em weighted voting games},
	which constitute a familiar example of voting systems.
In a weighted voting game, 
	each player has a voting weight 
	(intuitively corresponding to their contribution), 
	and a coalition wins
	if the sum of its members' weights meets or exceeds
	a given threshold.
	
Over the years, many power indices of voting games
	have been proposed
	and studied, such as the Shapley-Shubik index~\cite{
	shapley1954method}, 
	the Banzhaf index~\cite{banzhaf1964weighted},  
	and
	the Deegan-Packel index~\cite{deegan1978new}.
The problem of computing the players' power indices 
	has received ample attention 
	(see~\cite{matsui2000survey}	for example) 
	 and its computational
	complexity is well-understood~\cite{
	deng1994complexity,
	elkind2009computational,
	garey1979computers,
	matsui2001np,
	prasad1990np}.

The focus of this work is on the solution concept
	called the {\em least core}, proposed by 
	Maschler, Peleg, and Shapley~\cite{maschler1979geometric},
	which is a natural relaxation of the core.
Elkind et al.\@~\cite{elkind2009computational}
	showed some intractability results
	including coNP-hardness
	for checking the non-emptyness of 
	the $\varepsilon$-{\em core} of a weighted voting game.
They proposed a pseudo polynomial time algorithm
	to compute a payoff vector in the least core, 
	which is based on the (polynomial time) ellipsoid method 
	and a pseudo polynomial time separation oracle.
They also proposed 	
	a fully polynomial time approximation scheme (FPTAS) 
	for the value of the least core.
The {\em nucleolus} is a unique payoff vector
	with a lexicographically maximal excess vector,  
	which is contained in the least core
	(e.g., see~\cite{schmeidler1969nucleolus,elkind2009computational}). 
Pseudo-polynomial time algorithms for finding the nucleolus
	of a weighted voting game 
	are proposed 
	in ~\cite{elkind2009computing,pashkovich2018computing}.

In this paper, we propose 
	a pseudo-polynomial size LP ``formulation''
	for finding a payoff vector in the least core of a weighted voting game.
The numbers of variables and constraints
	of our formulation are both bounded by $\mbox{O}(n \sumW)$,
	where $n$ is the number of players and 
	$\sumW$ is the sum of (integer) voting weights.
Thus, a polynomial time algorithm
	for general linear programming problems
	finds a payoff vector in the least core in pseudo polynomial time.	
An advantage of our approach is that
	one can adopt his/her favored software (LP solver)
	for calculating a payoff vector in the least core.
The computational experiment shows the effectiveness and 
	efficiency of the proposed formulation. 		
In the last section, 
	we also extend our approach 
	to vector weighted voting games.

\section{Notations and Definitions}	
Let $N = \{1, 2,\ldots , n\}$ be a set of {\em players}. 
A weighted voting game $G$ is defined 
	by a sequence of positive integers
	$G=[q;w_1,w_2,\ldots ,w_n]$,
	where we may think of $w_i$ as the number of votes
	of player~$i$ 
	and $q$ as the quota 
	needed for a coalition to win. 
In this paper, we assume that 
	$0< q \leq w_1+w_2 + \cdots+ w_n$.
A coalition $S \subseteq N$ is called a {\em winning coalition}
	when the inequality $q \leq \sum_{i \in S}w_i $ holds. 
The set of all the winning coalitions is denoted by 
  $\cal W$.
%
A weighted voting game gives 
	a simple game $(N,v)$ defined by 
	a characteristic function $v:2^N \rightarrow \{0,1\}$ 
	ensuring that $v(S)=1$ if and only if $S \in {\cal W}$.

In this paper, we discuss a solution concept,
 	called the {\em least core},  
	of a characteristic function  game. 
Given a characteristic function game $(N, v)$, 
	a {\em pre-imputation} of $(N, v)$ is a non-negative payoff vector 
	$\Vec{x}=(x_i \mid i \in N)$ such that
	$\sum_{i \in N} x_i =v(N)$.
The {\em core} of $(N, v)$ is the set of pre-imputations
	satisfying $\sum_{i \in S}x_i \geq v(S)$ 
	for all $S \subseteq N$.
The $\varepsilon$-{\em core} of  $(N, v)$ 
	is the set of pre-imputations
	satisfying
	 $\sum_{i \in S}x_i \geq v(S)- \varepsilon$, 
	 $\forall S \subseteq N$. 
The {\em least core} of  $(N, v)$ is its $\varepsilon^*$-core
	where 
$
	\varepsilon^*= \inf \{ \varepsilon \mid 
		\varepsilon\mbox{-core of } (N, v) \mbox{ is non-empty}\}.
$

It is easy to see that 
	the least core of a weighted voting game
	is a set of vectors $\Vec{x} \in \Real^N$ 
	which confirm that $(\varepsilon^*, \Vec{x})$ is optimal to
	the following linear programming problem: 
\begin{alignat*}{2}
	\label{last}
		\mbox{P1: min. } \; & \varepsilon  && \\
		\mbox{s.t. } \; &\textstyle \sum_{i \in S} x_i \geq 1-\varepsilon 
			&\;\; &(\forall S \in {\cal W}), \\
		&\textstyle \sum_{i \in N}x_i = 1,\\
		&x_i\geq 0 && (\forall i \in N),
\end{alignat*}
where $\varepsilon^*$ is the optimal value of P1
  (see~\cite{chalkiadakis2011computational,elkind2009computational}).
Here we note that 
	the number of constraints of P1 may be exponential in $n$.

One of the most popular methods 
	to solve linear programming problems
	is the simplex method designed 
	by Dantzig~\cite{dantzig1963linear}.
It is quite efficient in practice and on average, 
	but no polynomial-time worst-case running time bound
	has been proved.
The ellipsoid method is 
	the first polynomial-time method for linear programming
	given by
	Khachiyan~\cite{khachiyan1980polynomial}.
The ellipsoid method is not competitive with the simplex method
	in practice (e.g., see \cite{bland1981ellipsoid}) 
	but it has important theoretical side-effects. 
Specifically, the ellipsoid method does not require
	listing all constraints
	of an LP problem a priori, 
	but allows that they are generated when needed. 
A {\em separation oracle} for a linear (feasibility) program 
	is a procedure that, given a candidate solution, 
	determines whether it is feasible and, if not, 
	outputs a violated constraint.
If a linear program 
	admits a separation oracle, 
	then an optimal solution can be found 
	by the ellipsoid method
	(see~\cite{grotschel2012geometric} for a detailed discussion).	
In 1984, Karmarkar~\cite{karmarkar1984new} 
	developed a method 
	for linear programming called Karmarkar's algorithm, 
	which runs in polynomial time.
Karmarkar's algorithm falls within the class of interior point methods
	characterized by polynomial complexity.
Moreover, 
	they have efficient implementations, 
	competing with the simplex method. 

When we solve P1 directly, 
	we need to generate all the (minimal) winning coalitions
	growing exponentially in $n$. 
Elkind et al.\@~\cite{elkind2009computational}
	adopted the ellipsoid method and proposed 
	a pseudo polynomial time separation oracle,
	which yields a pseudo polynomial time algorithm 
	for solving P1.

\section{Formulation} \label{sec:Formulation}

\begin{figure}[h]
	\centering
	\includegraphics[height=8cm]{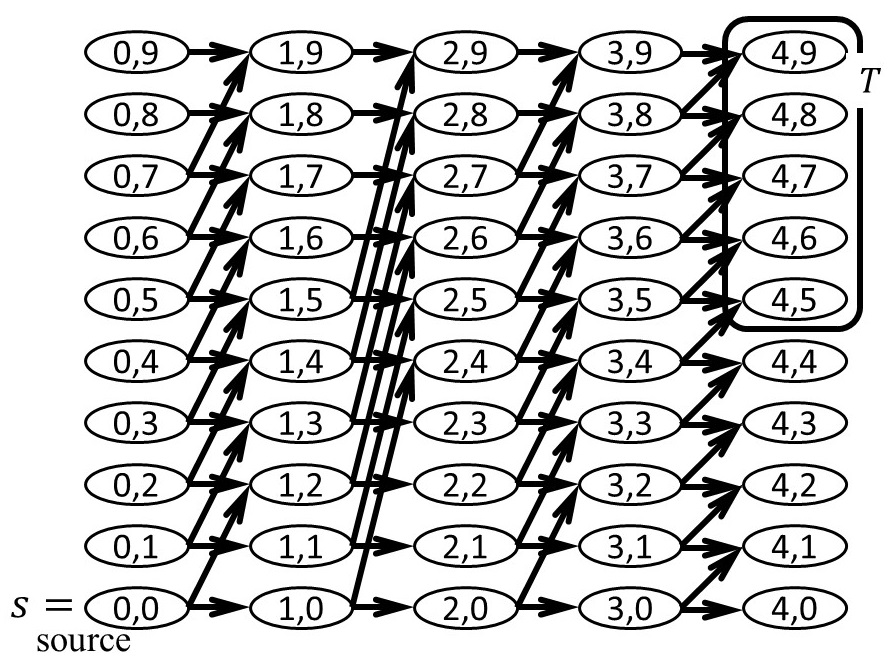}
	\caption{Directed graph $\Gamma$ corresponding to $G=[5; 2, 4, 2, 1]$.}			 						\label{fig:digraph}
\end{figure}

In this section, 
	we propose a new formulation for
	calculating a payoff vector in the least core of
	a weighted voting game. 
For any positive integer $\alpha$, 
  $[\alpha]$ and $[\alpha]_0$  denote the set of integers 
	$\{1,2,\ldots,  \alpha \}$ and $\{0,1,\ldots , \alpha \}$, respectively.
Here we note that the set of players $N$ is equal to $[n]$.
We define that $\sumW = \sum_{i=1}^n w_i$.
We introduce an acyclic directed graph
 $\Gamma =(V,A)$  with a vertex set  
 $V= [n]_0 \times [\sumW]_0$   
  and an arc set $A=A_0 \cup A_1 \cup \cdots \cup A_n$ 
	defined by
\[
\begin{array}{l}
	A_0=\{((i-1, \alpha), (i, \alpha)) 
		\mid (i, \alpha) \in [n] \times [\sumW]_0 \}, \\
	A_i=\{((i-1, \alpha), (i, \alpha+w_i)) 
		\mid \forall \alpha \in  [\sumW-w_i]_0 \} \;\;
		(\forall i \in N=[n]). 
\end{array}
\]
We define a specified set of vertices  
$
	T =\{(n, \alpha) \subseteq V \mid q \leq \alpha \}.
$
The vertex $(0,0) \in V$
	is called the {\em source} and denoted by $s$.
Figure~\ref{fig:digraph} shows an example of $\Gamma$.
A directed path in $\Gamma$
	from the source $s=(0,0)$ to a vertex in $T$
	is called an $s\hypn T$ path.
It is easy to see that 	
	there exists a bijection between 
	$\cal W$
	and a set of $s\hypn T$ paths.

Given a 
vector $\Vec{x} \in \Real^N$, 
	we introduce an arc length function 
	$w^{\Vec{x}}:A \rightarrow \Real$ 
	defined by
\[
		w^{\Vec{x}}(a)=
		\left\{
			\begin{array}{ll}
				0 	& (\mbox{if } a \in A_0), \\
				x_i & (\mbox{if } a \in A_i).
			\end{array}
 		\right.
\]
Then, it is easy to show that
	a pair $(\varepsilon, \Vec{x})$ 
 satisfies $\textstyle \sum_{i \in S} x_i \geq 1-\varepsilon 
			\;(\forall S \in {\cal W})$
	if and only if 
	the length (defined by $w^{\Vec{x}}$) 
	of the shortest $s\hypn T$ path 
	is greater than or equal to $1-\varepsilon$.
Given a pre-imputation $\Vec{x}$, 
	the length of the shortest $s\hypn T$ path, 
	denoted by $1-\varepsilon'$, satisfies 
	$\varepsilon'=\inf \{\varepsilon \mid 
		\varepsilon\mbox{-core of } (N, v) 
		\mbox{ includes } \Vec{x} \}$.
Thus, the problem of calculating the least core value 
	corresponds to a problem of finding a pre-imputation $\Vec{x}$ 
	which  maximizes the shortest  $s\hypn T$ path length.
Here we note that 
	we can transform the shortest $s\hypn T$ path 
	problem to an ordinary 2-terminal shortest path problem
	by contracting the set of vertices $T$ to 
	an artificial terminal vertex.

In the following part, 
	we employ a ``dual'' linear programming formulation
	for the shortest path problem.
An ordinary primal linear programming formulation for 
	the shortest path problem appears in~\mbox{\ref{PFSPP}}
	(otherwise, see~\cite{papadimitriou1981combinatorial}
	Section~5.4 for example).	
If we employ a primal formulation for 
	the shortest path problem,
	then the obtained formulation for the least core problem
	includes non-linear terms,
	which causes some difficulties
	(see~\mbox{\ref{PFSPP}} for details).

We introduce a vector of variables $(y(v) \mid v \in V)$
	and an artificial variable $y_T$.
Then, we have a linear programming formulation 
	of the shortest path problem: 
 \begin{alignat*}{2}
	\label{soutui}
		\mbox{D}(\Vec{x}): \; 
		{\rm {max.}}\ &y_T - y(s), &&\\
		{\rm s.t.} \;
		& y(v)-y(u) \leq 0 		&&(\forall (u, v)\in A_0), \\
		& y(v)-y(u) \leq x_i 	\:\:\:&& (\forall i \in N, \forall (u, v)\in A_i), \\
		& y(v) =y_T   &&(\forall v \in T).
\end{alignat*}
The last constraint $ y(v) =y_T  \; (\forall v \in T)$
	corresponds to a contraction operation 
	of  the set of vertices $T$ to 
	an artificial terminal vertex.


Since problem D$(\Vec{x})$ is a maximization problem,
	the length of the $s\hypn T$ shortest path
	is greater than or equal to $1-\varepsilon$
	if and only if  there exists a feasible solution of D$(\Vec{x})$
	satisfying $y_T - y(s) \geq 1-\varepsilon$.
Thus, $\Vec{x}\in \Real^N$ is in the least core
	if and only if  $\Vec{x}$ is a subvector 
	of an optimal solution to the following problem:
\begin{alignat*}{2}
		\mbox{P2: min. } \; & \varepsilon  && \\
		\mbox{s.t. } \; & y_T - y(s)\geq 1-\varepsilon, && \\
		& y(v)-y(u) \leq 0 	&&  (\forall (u,v)\in A_0), \\
		& y(v)-y(u) \leq x_i && (\forall i \in N, \forall (u,v)\in A_i),\\
		& y(v)=y_T  && (\forall v \in T), \\
		& \textstyle \sum_{i \in N}x_i = 1,\\
		&x_i\geq 0 && (\forall i \in N),
\end{alignat*}
where $\varepsilon$, $y_T$, $\{y(v) \mid v \in V\}$ 
	and $\{x_i \mid i \in N \}$ are continuous variables.
Here we note that 
	$\Vec{x}$ is a fixed vector in D$(\Vec{x})$ and
	a vector of variables in P2.

The numbers of variables and constraints
	in P2 are bounded by $\mbox{O}(n \sumW)$.
Thus, a polynomial time algorithm for general linear programming problems
	solves P2 and finds a payoff vector in the least core
	in pseudo polynomial time 
	on the input size of the integer vector 
	$[q;w_1,w_2,\ldots ,w_n]$.

In the following part, we summarize  known results 
	related to the behavior of $\sumW=\sum_{i =1}^n w_i$.
When a given simple game $(N, v)$ is a weighted voting game, 
	the corresponding integer-weights representation
	$[q; w_1, w_2, \ldots , w_n]$ is not unique.
Moreover, Isbell~\cite{isbell1956class} 
	found an example of a weighted simple game 
	with 12 players without a unique
	minimum-sum integer-weight representation.
Examples for 9, 10, or 11 players
	are given in
	Freixas and Molinero~\cite{freixas2009existence,freixas2010weighted}. 
The problem of determining a minimum-sum 
	integer-weight representation is computationally difficult, 
	as the problem of deciding a dummy player is 
	NP-complete~\cite{garey1979computers,	prasad1990np}	
	and a voting weight of every dummy player is equal to zero
	in a  minimum-sum integer-weight representation.
In cases of $n \leq 9$ players,
	 Kurz~\cite{kurz2012minimum} enumerated 
	 (minimal) integer-weight representations.
In general, Kawana and Matsui~\cite{kawana2021trading}
	showed that 
	every weighted voting game has an integer-weight representation
	satisfying $\sumW \leq (n+2)^{\frac{n+2}{2}}(1/2)^n -1$.
In the next section, we describe two practical examples 
	satisfying $(n, \sumW)=(51, 538), (27, 345)$.
	

\section{Computational Experiences} \label{section:CompExp}

This section reports the results 
	of our 
	numerical experiments.
%
All the experiments were conducted on a PC running 
	the Windows 10 Pro operating system 
	with an Intel(R) Core(TM) i7-8700 @3.2GHz processor and 
	16 GB RAM.
All instances were solved using 
	CPLEX 12.8.0.0 implemented in Python 3.5.4.

A weighted voting game reported in~\cite{owen1995game}
(Section~12.4)
	for a voting process in the United States
	has a quota and a vector of weights
\[
\begin{array}{l}
[270; 
	45, 
	41, 
	27, 
	26, 26, 
	25, 
	21, 
	17, 17, 
	14, 
	13, 13, 
	12, 12, 12, 
	11,
	\underbrace{10, \ldots , 10}_{\mbox{4 times}},
	\underbrace{9, \ldots , 9}_{\mbox{4 times}}, 
\\
	8, 8, 
	\underbrace{7, \ldots , 7}_{\mbox{4 times}},
	\underbrace{6, \ldots , 6}_{\mbox{4 times}},
	5,
	\underbrace{4, \ldots , 4}_{\mbox{9 times}},
	\underbrace{3, \ldots , 3}_{\mbox{7 times}}
],
\mbox{ where } n=51 \mbox{ and } \sumW=538.
\end{array}
\]
We used the above data and solved problem P2.
The computational time is 1.343 seconds and
	the optimal value $\varepsilon^*$ is equal to $0.49814\cdots$.
The  obtained vector in the least core is proportional
	to the vector of voting weights to ensure that
	the sum is equal to 1.

A game of the power of countries in the EU Council~\cite{
	felsenthal2001treaty,
	bilbao2002voting} 
	with 27 players is defined by
\[
\begin{array}{r}
	[255; 29, 29, 29, 29, 27, 27, 14, 13, 12, 12, 12, 12, 12, 
	10, 10, 10, \qquad \\
	7, 7, 7, 7, 7, 4, 4, 4, 4, 4, 3],
	\mbox{ where } n=27 \mbox{ and } \sumW=345. 
\end{array}
\]
In this case, the computational time for solving P2
	is 0.063 seconds and the optimal value $\varepsilon^*$ 
	is equal to $0.26086\cdots$.
As in the case of the voting process in the United States, 
	the  obtained vector in the least core is proportional 
	to the vector of voting weights.
Here we note that a vector in the least core 
	is not necessarily unique, and thus 
	vectors ``obtained'' by solving P2 are proportional to
	the voting weights in the above two cases.

We generated random artificial instances as follows.
For each pair $(n, U) \in \{40, 60, 80, 100, 120\}^2$, 
	we constructed $20$ weighted voting games 
	by generating voting weight $w_i\; (i \in \{1,2,\ldots , n\})$ 
	uniformly at random from 
	the set of integers $\{1,2, \ldots, U\}$.
We set $q$ to an integer obtained by rounding
	 $0.9 \sumW=0.9 \sum_{i=1}^n w_i$.
Table~\ref{addcomp} shows the average of computation time 
	of 20 instances generated for each cell. 
We applied the least-squares  regression method 
	to $5\cdot 5 \cdot 20=500$ instances.
The estimated regression equation is as follows: 
\[
\ln (\mbox{comp. time [s]})
	= -25.208+2.885 \ln n + 1.891 \ln \sumW. 
\]

\begin{table}[hbt]
\begin{center}
\caption{Computation time [s].} \label{addcomp}

\begin{tabular}{c|rrrrr}
$U$ 
& \multicolumn{5}{c}{$n$  (number of players)} \\
\cline{2-6}
(\mbox{UB of }$w_i$)  & 40 & 60 & 80 & 100 & 120 \\
\hline
40 & 0.2 & 1.1 & 3.6 & 9.7 & 20.4 \\
60 & 0.4 & 2.3 & 9.2 & 23.0 & 47.9 \\
80 & 0.5 & 3.9 & 14.7 & 42.9 & 91.9 \\
100 & 0.6 & 5.5 & 23.9 & 68.7 & 198.5 \\
120 & 0.8 & 7.6 & 35.6 & 121.1 & 280.7 \\
\end{tabular}
\end{center}
\end{table}

We give an additional experiment for checking 
	whether a vector in the least core is proportional to 
	a given vector of voting weights.
In general, this property does not hold.
We need only recall examples of dummy players 
	with positive voting weights.
Our experiences are limited to
	randomly generated artificial problems
	with small numbers of players.
We set the number of players at  $n \in \{8, 10, 12, 15, 18\}$
	and generated 100 vectors of weights for each $n$.
A voting weight $w_i$ for each player $i$ is chosen
	uniformly at random from the set of integers $\{1,2, \ldots , 20\}$.
For each generated vector of weights, 
	we tested all the cases 
	in which a quota $q$ is equal to an integer
	satisfying $(1/4)\sumW \leq q \leq (3/4) \sumW$.
Table~\ref{result} shows the results of our experiment.
The rows denoted by ``\#  VLC $ \parallel $ weights''
	(and ``\# VLC $\not \, \parallel $ weights'') 
	show the number of instances
	satisfying the condition 
	that the obtained vector in the least core
	is (not) proportional to the voting weights, respectively.
The row denoted by ``ratio [\%]'' shows the ratio of 
	the number of instances in the  row ``\#  VLC $ \parallel $ weights''
	to the total number of generated instances.
When $n=18$, all the generated instances satisfy
	the condition that
	the obtained vector in the least core is proportional to 
	the voting weights.  
Our results indicate a property of the nucleolus 
	of weighted voting games shown by
	Kurz, Napel, and Nohn~\cite{kurz2014nucleolus},
	who proved that under some assumptions, 
	when the relative weight of every player tends to zero, 
	the nucleolus converges to a vector 
	proportional to the voting weights.

\begin{table}[h]
	\begin{center}
		\caption{A payoff vector in the least core and voting weights.}

		\begin{tabular}{r|rrrrrr} 
		\# of players &5 & 8& 10 & 12 & 15 & 18\\ \hline
		\#  VLC $ \parallel $ weights
		 &46&89&1088&4182&7251 & 9060\\ 
		\# VLC $\not \, \parallel $ weights &2490&3792&3987&1833&167 & 0\\ \hline
		ratio [\%]&1.81&2.29&21.44 &69.53 & 97.75 & 100\\ 
		\end{tabular}
		\label{result}
	\end{center}
\end{table}

\section{Vector Weighted Voting Games}

In this section, 
	we discuss a case of vector weighted voting games, briefly.
Chalkiadakis, Elkind, and Wooldridge~\cite{chalkiadakis2011computational}
	summarized previous works related 
	to vector weighted voting games.

Let $k$ be a given positive integer.
For any $k' \in \{1,2,\ldots , k\}$, 
	let
	$G_{k'}=[q_{k'}; w_{k'1}, w_{k'2}, \ldots , w_{k'n}]$
	be a weighted  voting game
	defined on a mutual set of players $N=\{1, 2, \ldots , n\}$.
The corresponding sets of winning coalitions are denoted by 
	${\cal W}_{k'}$  $(k'\in \{1,2,\ldots ,k\})$.
An intersection of $k$ games $G_1, G_2, \ldots , G_k$ is 
	a simple game $(N, v_{\wedge})$ defined by 
	a characteristic function $v_{\wedge}:2^N \rightarrow \{0,1\}$ 
	satisfying  $v_{\wedge}(S)=1$
	 if and only if  $S \in {\cal W}_{k'}$ $(\forall k' \in \{1,2, \ldots , k\})$.
The dimensionality, $k$, of a $k$-vector weighted voting game
	can be understood as a measure of the complexity 
 	of the game. 
Elkind et al.\@~\cite{elkind2008dimensionality} 
	showed that the problems of `equivalence'
	and `minimality' of $k$-vector weighted voting games
	are computationally difficult.
They also provided efficient algorithms
	for those cases where two conditions are met:
	$k$ is small 
	and the weights are polynomially bounded. 
                               
For calculating a payoff vector
	in the least core of $(N, v_{\wedge})$,
	we only need to change the definition 
	of the directed graph 
	$\Gamma=(V, A)$ as follows.
We introduce a vertex set 
	$V=[n]_0 \times [\sumWk{1}]_0 \times \cdots \times [\sumWk{k}]_0$
	where $\sumWk{k'}=\sum_{i=1}^n w_{k'i}$ $(k' \in \{1, 2, \ldots , k\})$.
A vertex in $V$ is denoted by 
	$(i,\Vec{\alpha})=(i,\alpha_1,\alpha_2, \ldots ,\alpha_k) \in \Int^{k+1}$.
For each player $i \in N$, 
	we denote the vector $(w_{1i},w_{2i}, \ldots , w_{ki})$ by 
	$\Vec{w_{*i}}$.
An arc set $A=A_0 \cup A_1 \cup \cdots \cup A_n$ is
	defined by
\[
\begin{array}{l}
	A_0=\{((i-1, \Vec{\alpha}), (i, \Vec{\alpha})) 
		\mid i \in [n], \Vec{\alpha} 
			\in [\sumWk{1}]_0 \times \cdots \times [\sumWk{k}]_0 \}, \\
	A_i=\left\{
		((i-1, \Vec{\alpha}), (i, \Vec{\alpha}+\Vec{w_{*i}}))
	 \left| 
	 	\begin{array}{l}
	 		(i-1,\Vec{\alpha}) \in V \mbox{ and } 
	 		(i, \Vec{\alpha}+\Vec{w_{*i}}) \in V 
	 	\end{array}
	 \right. \right\}, 	
\end{array}
\]
for each $i \in N= [n]$.
We also set 
$
	T =\{(n, \Vec{\alpha}) \subseteq V 
		\mid q_{k'} \leq \alpha_{k'} \;\; (\forall k' \in \{1,2, \ldots ,k\})   \}.
$
Then, it is easy to see that problem P2 finds 
	a payoff vector in the least core of a simple game $(N, v_{\wedge})$.
The numbers of variables and constraints of P2	
	are bounded by $O(n \prod_{k'=1}^k \sumWk{k'})$
	and thus P2 remains a pseudo polynomial of size LP.

A union of $k$ games $G_1, G_2, \ldots ,G_k$ is 
	a simple game $(N, v_{\vee})$ defined by 
	a characteristic function $v_{\vee}:2^N \rightarrow \{0,1\}$ 
	satisfying $v_{\vee}(S)=1$
	 if and only if  $\exists k' \in \{1,2, \ldots , k\}, S \in {\cal W}_{k'}$.
We can deal with this case by setting
$
	T =\{(n, \Vec{\alpha}) \subseteq V 
		\mid \exists k' \in \{1,2, \ldots , k\},
		q_{k'} \leq \alpha_{k'}  \}.
$

It is easy to see that 
	we can extend our formulation 
	to a Boolean weighted voting game~\cite{faliszewski2009boolean,kurz2016dimension}
	obtained by combining some weighted voting games
	according to a complex Boolean formula.
	
\section{Conclusion} \label{conclusion}

In this paper, 
	we proposed a pseudo polynomial size 
	linear programming formulation 
	for calculating a payoff vector in the least core
	of a weighted voting game.
When we employ our formulation,
	a commercial solver finds a payoff vector in the least core 
	of practical weighted voting games
	in a few seconds.
Our computational experiences indicate that
	a vector in the least core is proportional to
	the voting weights in many cases. 

We also extended our formulation to vector weighted 
	voting games.
Further research is needed
	to evaluate the computational performance
	of the formulation for vector weighted voting games.

\appendix

\section{Primal Formulation of Shortest Path Problem}
\label{PFSPP}

In this section, 
	we describe an ordinary primal formulation 
	of the  $s\hypn T$ shortest path defined on 
	 the acyclic directed graph $\Gamma =(V, A)$
	 with arc length function 
	 $w^{\Vec{x}}:A \rightarrow \Real$.  
First, we transform the problem 
	into an ordinary 2-terminal shortest path problem
	by introducing an artificial terminal vertex $\tT$
	and artificial arcs $\tArc =T \times \{\tT\}$ 
	with $0$-length.
We denote the obtained  acyclic directed graph
	by $\tGamma =(V \cup \{ \tT \}, A \cup \tArc)$. 
For each vertex $v \in V \cup \{ \tT \}$, 
	we denote by $\delta^+ (v)$ and $\delta^- (v)$
	the set of outgoing and incoming arcs 
	of vertex $v$ in $\tGamma$.
For each arc $a \in 	 A \cup \tArc$, 
	we introduce a variable $\xi (a)$
	that takes the value $1$ 
	if the arc $a$ belongs to a shortest path.
A standard primal linear programming formulation 
	to determine a shortest path
	from $s$ to $\tT$ on $\tGamma$ is the following:
\begin{alignat*}{2}
		\mbox{P}(\Vec{x}): \; 
		{\rm {min.}}\ & \sum_{i=1}^n \sum_{a \in A_i} x_i \xi (a), &&\\
		{\rm s.t.} \;
		& \sum_{a \in \delta^+ (v)} \xi (a)
		 - \sum_{a \in \delta^- (v)} \xi (a)
		 = \left\{ \begin{array}{l}
		 	1 \\ -1 \\ 0 
		 		\end{array} \right. &&
		 		\begin{array}{l}
		 			(v=s), \\ (v=\tT), \\(\mbox{otherwise}),
				\end{array}		 		\\
	&\xi(a) \geq 0 		&&(\forall a \in A). \\
\end{alignat*}
Problem D$(\Vec{x})$ defined in Section~\ref{sec:Formulation},
	is obtained from the dual of P$(\Vec{x})$
	by  trivial transformations.

By setting the length of the $s\hypn \tT$ shortest path 
	on $\tGamma$ to $1-\varepsilon$,
 	$\Vec{x}\in \Real^N$ is in the least core
	if and only if  $\Vec{x}$ is a subvector 
	of an optimal solution to the following problem: 
\begin{alignat*}{2}
		\mbox{P': min. } \; & \varepsilon  && \\
		\mbox{s.t. } \; & 1-\varepsilon = z(\Vec{x})  && \\
		& \textstyle \sum_{i \in N}x_i = 1,\\
		&x_i\geq 0 && (\forall i \in N),
\end{alignat*}	
	where $z(\Vec{x})$ denotes the optimal value 
	of P$(\Vec{x})$.
The above formulation  implicitly includes
	nonlinear terms  $x_i \xi (a)$, 
	which are difficult to address directly.
The problem P' corresponds to 
	the shortest path problem with variable arc length,
	where related problems are discussed 
	in~\cite{magnanti1990shortest,cappanera2011optimal}
	for example.



\bibliographystyle{elsarticle-num} 
\bibliography{LeastCoreRefer4}





\end{document}